\documentclass[aps,prl,10pt,twocolumn,altaffiliation]{revtex4-2}

\usepackage{times}
\usepackage{changes}
\usepackage{graphicx}
\usepackage{amssymb}
\usepackage{epstopdf}
\usepackage{amsmath}
\usepackage{color}
\usepackage{hyperref}
\usepackage{bbold}
\usepackage{bm}
\usepackage{enumitem}
\usepackage[acronym]{glossaries}
\usepackage{physics}
\makeatletter
\graphicspath{{Images/}}

\usepackage{hyperref}
\hypersetup{
	colorlinks=true,
	linkcolor=black,          
	citecolor=blue,           
	filecolor=magenta,        
	urlcolor=cyan
}

\newcommand{\eqdef}{\mathrel{\mathop:}=}
\newcommand{\pprime}{{\prime\prime}}

\makeatother	

\begin{document}
	
	\title{Quantum Trails and Memory Effects in the Phase Space of Chaotic Quantum Systems}
	
	\author{Andrea Pizzi}
	\email{ap2076@cam.ac.uk}
	\affiliation{Cavendish Laboratory, University of Cambridge, Cambridge CB3 0HE, United Kingdom}
	
	\begin{abstract}
		The eigenstates of a chaotic system can be enhanced along underlying unstable periodic orbits in so-called quantum scars, making it more likely for a particle launched along one such orbits to be found still there at long times. Unstable periodic orbits are, however, a negligible part of the phase space, and a question arises regarding the structure of the wave function elsewhere. Here, we address this question and show that a weakly-dispersing dynamics of a localized wave packet in phase space leaves a ``quantum trail'' on the eigenstates, that is, makes them vary slowly when moving along trajectories in phase space, even if not periodic. The quantum trails underpin a remarkable dynamical effect: for a system initialized in a localized wave packet, the long-time phase-space distribution is enhanced along the short-time trajectory, which can result in ergodicity breaking. We provide the general intuition for these effects and prove them in the stadium billiard, for which an unwarping procedure allows us to visualize the phase space on the two-dimensional space of the page.
	\end{abstract}
	
	\maketitle
	Classical chaos is famously characterized by the butterfly effect, where small differences in initial conditions lead to vastly different outcomes over time~\cite{lorenz1963deterministic,strogatz2018nonlinear,wimberger2014nonlinear}. If and how chaos manifests in quantum systems is a more nuanced question, the center of the mature field of quantum chaos~\cite{berry1989quantum,jensen1992quantum,delande1994classical,wimberger2014nonlinear}. On the surface, strongly chaotic quantum systems behave similarly to random matrices: their level statistics follows random matrix theory according to the BGS conjecture~\cite{bohigas1984characterization,berry1987quantum,atas2013distribution,kos2018many}, and their eigenstates appear rather featureless~\cite{berry1977regular,hortikar1998correlations,rigol2008thermalization}. But, of course, physical systems are not random matrices, and deviations can occur. A striking example are quantum scars, whereby the eigenstates can show enhanced amplitude along certain classical unstable periodic orbits~\cite{heller1984bound}. This enhancement means that a quantum particle is more likely to remain localized along an orbit it was prepared on, which can lead to a form of ergodicity breaking~\cite{kaplan1998linear}. More recently, these phenomena have been shown to play an important role in our understanding of isolated many-body systems out of equilibrium~\cite{turner2018weak,serbyn2021quantum,pilatowsky2021ubiquitous,hummel2023genuine,ermakov2024periodic,pizzi2024quantum,evrard2024quantuma,evrard2024quantumb}, as increasingly relevant in modern quantum simulators~\cite{gross2017quantum,bernien2017probing,li2023improving}.
	
	Yet, unstable periodic orbits occupy only an infinitesimally small fraction of the phase space, raising a natural question: how does the quantum wave function behave in the vast regions of phase space that are not tied to these special orbits?
	
	Here, we address this question and find that, if a wave packet moves along a trajectory with little dispersion, then a ``quantum trail'' is left on the eigenstates, namely, the projection of the eigenstates in phase space varies slowly along the trajectory. As a direct consequence of quantum trails, an initially localized wave packet fails to uniformly scramble across the accessible phase space: the long-time distribution is enhanced along the short-time trajectory, a memory effect that can break ergodicity. We illustrate these concepts in the paradigmatic stadium billiard, for which a phase-space unwarping procedure facilitates the visualization of the trails. Our work unveils the structure of the eigenstates in phase space, shows its implications on the long-time dynamics, and contributes to our understanding of the classical-quantum correspondence.
	
	\begin{figure}
		\centering
		\includegraphics[width=\linewidth]{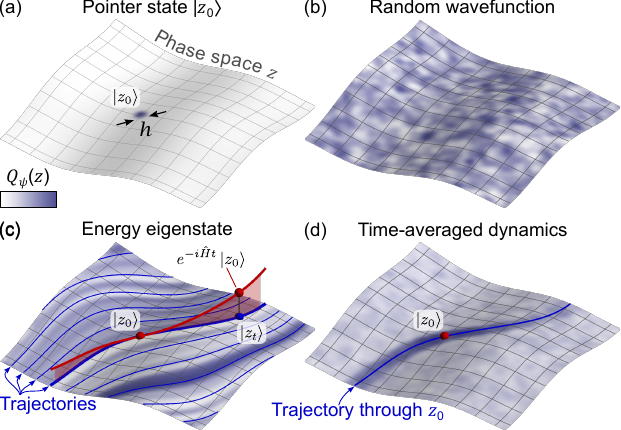}
		\caption{\textbf{Quantum states in phase space: trails and memory effects.} Schematic projections $Q_\psi(z)$ on a phase space of coordinates $z$.
			(a) A pointer state $\ket{z_0}$ is localized around $z_0$ over some length $\sim h$.
			(b) A random wave function yields a speckled pattern with features of size $\sim h$.
			(c) An eigenstate $\ket{E}$ can yield \textit{quantum trails}, that is, speckles of width $\sim h$ and length $> h$ that elongate along the trajectories $z_t$ (blue lines). The trails extend as long as $\ket{z_t} \approx e^{-i \hat{H} t} \ket{z_0}$.
			(d) Because of the trails, a system initialized in $\ket{z_0}$ is, at long times, more likely to be found on the short-time trajectory through $z_0$, a memory effect that can break ergodicity. }
		\label{Fig1}
	\end{figure}
	
	\textit{General intuition and phenomenology}---Consider a quantum system with Hamiltonian $\hat{H}$ and a phase space of coordinates $z = (z_1, z_2, \dots)$. Denote $\ket{z}$ a pointer state localized around the point $z$ of the phase space, Fig.~\ref{Fig1}(a), and say $h$ the characteristic localization length. The projection of a wave function $\ket{\psi}$ on phase space reads $Q_{\psi}(z) = \left| \langle z | \psi \rangle \right|^2$, and that of a random (e.g., Haar random) state $\ket{\psi_{\rm rand}}$ appears as a chaotic speckled pattern with correlation length $\sim h$, Fig.~\ref{Fig1}(b).
	
	The eigenstates $\ket{E}$ can be strikingly different. The key point is: if the pointer state $\ket{z_t}$ associated to a trajectory $z_t$ in phase space is close to the actual dynamics from $\ket{z_0}$, namely, if $e^{-i \hat{H} t}\ket{z_0} \approx \ket{z_t}$, then it immediately follows that $\langle E | z_t\rangle \approx e^{-iEt} \langle E | z_0\rangle$ and $Q_E(z_t) \approx Q_E(z_0)$, even if $|z_t - z_0| \gg h$. 
	In other words, if wave packets in phase space evolve without immediately dispersing, $e^{-i \hat{H} t}\ket{z_0} \approx \ket{z_t}$, then the eigenstates in phase space cannot consist of a speckled pattern as in Fig.~\ref{Fig1}(b), but must instead consist of \textit{elongated} speckles, or ``\textit{quantum trails}'', as in Fig.~\ref{Fig1}(c). The trails imply a rich correlation structure in phase space: while the projections $Q_E(z_0)$ and $Q_E(z_t)$ can behave like pseudorandom numbers (e.g., with Porter-Thomas distribution~\cite{porter1956fluctuations,boixo2018characterizing}), they will be correlated with each other if $z_t$ is on the short-time trajectory through $z_0$. The ``short-time trajectory'' is the segment of trajectory for which the condition $\ket{z_t} \approx e^{-i \hat{H} t}\ket{z_0}$ still holds. Its length determines that of the trails and of the correlations in $Q_E(z)$. In contrast to scars~\cite{kaplan1998linear}, the trails do not require unstable periodic orbits nor localization.
	
	The structure of the eigenstates has direct and dramatic effects on the dynamics. The question we ask is: starting from $\ket{z_0}$, where does the system end up at long times, on average? This is quantified by the time-averaged phase-space projection~\cite{nordholm1974quantum,heller1987quantum,heller1980quantum,kaplan2000short,heller2018semiclassical,pizzi2024quantum}
	\begin{align}
		\bar{Q}(z|z_0)
		& = \lim\limits_{T \to \infty} \frac{1}{T} \int_0^T dt \
		\left| \langle z| e^{-i \hat{H} t} | z_0 \rangle \right|^2, \\
		& = \sum_E Q_E(z_0) Q_E(z),
		\label{eq. Qbar}
	\end{align}
	which we expanded in the energy basis assuming a nondegenerate spectrum. For an eigenstate $\ket{E}$ to significantly contribute to the sum, both $Q_E(z_0)$ and $Q_E(z)$ should be relatively large, that is, the initial condition $\ket{z_0}$ should have a significant component over the eigenstate and the eigenstate should matter for the observation point $z$. A correlation between $Q_E(z_0)$ and $Q_E(z)$ means that, if one condition is met, the other is met ``for free'', thus enhancing $\bar{Q}(z|z_0)$. The extreme case is $z = z_0$, for which the correlation is perfect and the enhancement of $\bar{Q}(z_0|z_0)$ known~\cite{heller1980quantum,kaplan2000short,heller2018semiclassical}.
	
	Crucially, the quantum trails in the eigenstates imply that the correlation between $Q_E(z_0)$ and $Q_E(z)$ stretches along the short-time trajectory $z_t$ through $z_0$, in turn implying an enhanced time-averaged projection $\bar{Q}$ there, see Fig.~\ref{Fig1}(d). One could say that quantum mechanics leaves no second chances: if at short times the system does not quickly disperse in phase space, it never fully will. The effect is particularly remarkable if the underlying trajectories are ergodic: a classical ensemble fills the phase space uniformly at long times, but a quantum wave packet does not, retaining information on the initial condition and thus breaking ergodicity~\cite{heller1980quantum}. Under standard assumptions for chaotic quantum systems, in Appendix A we estimate the contrast of the enhancement in $\bar{Q}$, namely, the ratio between $\bar{Q}(z_t|z_0)$ on a point $z_t$ of the trajectory through $z_0$ and $\bar{Q}(z_g|z_0)$ on a generic point $z_g$ of the accessible phase space,
	\begin{equation}
		\frac{\bar{Q}(z_t|z_0)}{\bar{Q}(z_g|z_0)}
		\approx 1 + \left| \langle z_t | e^{-i \hat{H} t} | z_0 \rangle \right|^2.
		\label{eq. contrast}
	\end{equation}
	The contrast is $\approx 2$ when $\ket{z_t} \approx e^{-i \hat{H} t} \ket{z_0}$, suggesting that in the classical limit $h \to 0$ the enhancement in $\bar{Q}$ becomes infinitesimally narrow but does not loose contrast, and recovering the doubled return probability in the limit cases of $\ket{z_t} = e^{-i \hat{H} t} \ket{z_0}$ and $t=0$~\cite{heller2018semiclassical}.
	
	The treatment was so far general and prioritized conceptual clarity over technical precision. We have deliberately kept the notion of phase space $z$, trajectories $z_t$, and pointer states $\ket{z}$ vague. Their choice is indeed ultimately arbitrary and problem dependent. For a semiclassical system, $z$ and $z_t$ can be naturally taken from the classical limit. For a many-body system, $z$ could be the parameters of a variational wave function $\ket{z}$, and the trajectories $z_t$ given by the time-dependent variational principle (TDVP)~\cite{haegeman2011time,haegeman2016unifying,ho2019periodic,michailidis2020slow}. In the end, the key condition for the quantum trails is that $\ket{z_t}$ is for some time a good approximation of the actual dynamics, namely, that $\left| \langle z_t | e^{-i \hat{H} t} | z_0 \rangle \right|^2 \sim 1$, and this should drive the choice of $z$, $z_t$, and $\ket{z}$ for a given $\hat{H}$.
	
	\begin{figure*}
		\centering
		\includegraphics[width=\linewidth]{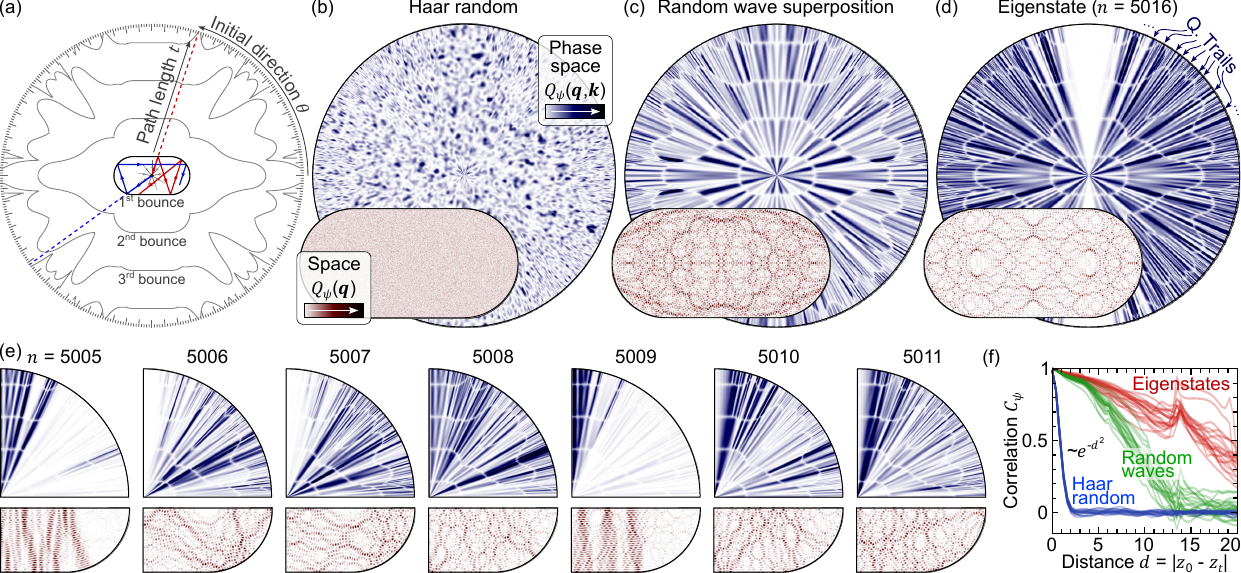}
		\caption{\textbf{Quantum trails in a chaotic billiard.}
			(a) An unwarping technique allows us to represent the three-dimensional phase space on the two-dimensional page, as follows. Consider a trajectory $(\bm{q}_t, \bm{k}_t)_\theta$ (solid red) from $\bm{q}_0 = (0,0)$ and $\bm{k}_0 = k(\cos \theta, \sin \theta)$. We associate the points $\bm{q} = \bm{k}_0 t$ of the page (red dashed line) to the points $(\bm{q}_t, \bm{k}_t)_\theta$ of the phase space. The polar coordinates $t$ and $\theta$ on the page can be used to move in phase space along the trajectories and away from them, respectively.
			(b) A Haar random state yields a speckled pattern in phase space (blue), and a collection of random pixels in space (red).
			(c) A random wave superposition yields, in phase space, partial trails that extend along trajectories but are interrupted by the collisions with the boundary of the billiard.
			(d) An eigenstate $\ket{k_n}$ is similar to a random wave superposition in space, but strikingly different in phase space: full-fledged quantum trails stretch along the trajectories (radially on the page), even across collisions with the boundary.
			(e) Projection of consecutive eigenstates $\ket{k_n}$. Some eigenstates are scarred, namely, localized along unstable periodic orbits (e.g., $n = 5005$ and $5009$). The quantum trails are a more general feature: the eigenstates (both scarred and not) correlate along the trajectories (both periodic and not).
			(f) Pearson correlation coefficient between $Q_\psi(z_0)$ and $Q_\psi(z_t)$ versus distance $d = |z_0 - z_t|$, computed sampling points $z_0$ uniformly in phase space and points $z_t$ along the short-time ($t<3/\lambda \approx 3.49/k$) dynamics from $z_0$. For Haar random states (blue), the correlation trivially decays as $\sim e^{-d^2}$. For the random wave superpositions (green), but even more for the eigenstates (from $n = 5009$ to $n = 5028$, red), the correlation extends over large distances $d$, due to the quantum trails. Here, $k = 94.68$.}
		\label{Fig2}
	\end{figure*}
	
	\textit{The stadium billiard}---We prove these ideas in a paradigmatic model of quantum chaos: the stadium billiard~\cite{bunimovich1974ergodic, bunimovich1979ergodic}. This consists of a particle in a box, $\hat{H} = \frac{\bm{\hat{p}}^2}{2m} + V(\hat{\bm{q}})$, with $V(\bm{q}) = 0$ inside the billiard and $V(\bm{q}) = \infty$ outside of it. We set $m = \hbar = 1$ and consider the billiard of height $2R = 2$ and width $2R + L = 4$. We use the boundary integral method to numerically find the spectrum $\hat{H} \ket{k_n} = \frac{k_n^2}{2} \ket{k_n}$~\cite{backer2003numerical,SM}.
	
	The phase space consists of position $\bm{q} = (q_x,q_y)$ and momentum $\bm{k} = (k_x,k_y)$. The trajectories $z_t$ are the classical ones and the momentum $\bm{k} = k(\cos \theta, \sin \theta)$ has conserved $|\bm{k}| = k$, so that the phase space can be reduced to three dimensions, $z = (q_x,q_y,\theta)$. As pointer states $\ket{\bm{q}\bm{k}}$ we consider Gaussian wave packets with centroid $(\bm{q},\bm{k})$, which in the position basis read $\langle \bm{q}^\prime | \bm{q}\bm{k} \rangle = \frac{1}{\sigma \sqrt{\pi}} e^{-\frac{|\bm{q} - \bm{q}^\prime|^2}{2 \sigma^2} + i \bm{q}^\prime \cdot \bm{k}}$~\cite{bohigas1993manifestations,heller2018semiclassical,sakurai2020modern}. The respective uncertainties $\langle (\Delta q_{x,y})^2 \rangle = \frac{\sigma^2}{2}$ and $\langle (\Delta k_{x,y})^2 \rangle = \frac{1}{2 \sigma^2}$ saturate the Heisenberg uncertainty principle, and to ensure a similar localization in position and momentum we enforce $\frac{\langle (\Delta q_{x,y})^2 \rangle}{R^2} = \frac{\langle (\Delta k_{x,y})^2 \rangle}{k^2} \ll 1$, that is, set $\sigma^2 = \frac{1}{k}$ and work in the semiclassical regime $k \gg 1$~\cite{SM}.
	
	A wave function $\ket{\psi}$ can be visualized using either the space projection $Q_{\psi}(\bm{q}) = \left| \langle \bm{q} | \psi \rangle \right|^2$ or the phase-space projection $Q_{\psi}(\bm{q}\bm{k}) = \left| \langle \bm{q}\bm{k} | \psi \rangle \right|^2$. The former is the standard probability density in space and allows us to view scars~\cite{heller1984bound}, whereas the latter corresponds to $Q_\psi(z)$ in our general formalism above and allows us to view the quantum trails. Visualization in phase space is in fact nontrivial, due to its dimensionality $>2$. Previous work has focused on a Poincar\'{e} section by rewriting the dynamics $z_t$ as a map between Birkhoff coordinates at consecutive bounces~\cite{crespi1993quantum, tualle1995normal, simonotti1997quantitative, cerruti2000exploring, backer2004poincare}. This technique however does not allow a clear visualization of the quantum trails, because the trajectories appear in the Poincar\'{e} section as sequences of disconnected points, breaking the trails into pieces~\cite{schanz2005phase,harayama2015ray}. We therefore devise an alternative strategy to represent the three-dimensional phase space on the two-dimensional page.
	
	The task of scanning the phase space can in fact be effectively delegated to the trajectories: due to ergodicity~\cite{bunimovich1974ergodic, bunimovich1979ergodic}, even a single ``probe'' trajectory $(\bm{q}_t, \bm{k}_t)$ will at $t \to \infty$ explore the whole phase space (excluding the measure-0 set of periodic orbits). That is, the finite three-dimensional phase space of coordinates $z = (q_x, q_y, \theta)$ can be ``unwarped'' onto an infinite one-dimensional space of parametric coordinate $t$, allowing us to inspect how the projection $Q_\psi(\bm{q}_t \bm{k}_t)$ varies along a trajectory. To fully appreciate a quantum trail, however, we also need to inspect the phase space in a direction orthogonal to $z_t$. Thus, instead of launching a single probe trajectory we launch a fan of them, $(\bm{q}_t,\bm{k}_t)_{\theta}$, starting from the middle of the billiard with direction $\theta$ varying continuously in $[0,2\pi]$, see gray lines in the middle of Fig.~\ref{Fig2}(a). A point of the page with polar coordinates $\theta$ and $t$ will then correspond to a point of the phase space $(\bm{q}_t, \bm{k}_t)_\theta$. Varying $t$ and $\theta$ allows moving along trajectories or away from them, respectively. In this representation, the quantum trails correspond to features that stretch radially on the page.
	
	\begin{figure*}
		\centering
		\includegraphics[width=\linewidth]{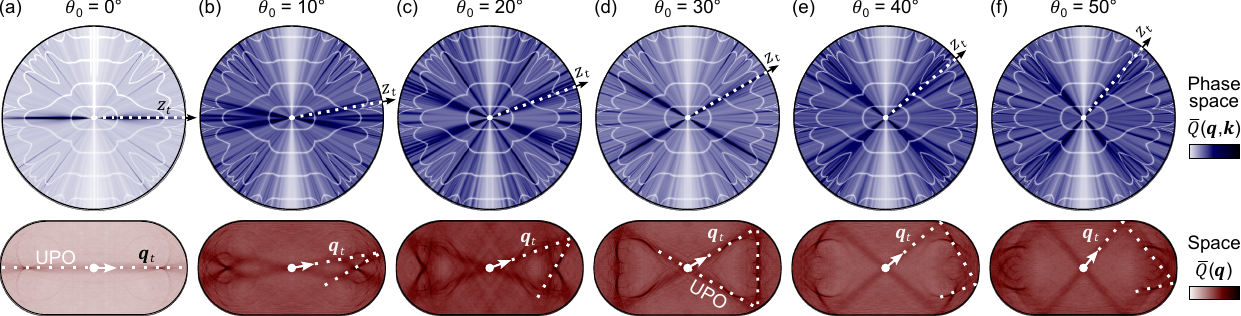}
		\caption{\textbf{Memory effect and ergodicity breaking in a chaotic billiard.} (a-f) We solve the dynamics $e^{-i \hat{H} t} \ket{\bm{q}_0\bm{k}_0}$ from a wave packet $\ket{\bm{q}_0\bm{k}_0}$ launched from the middle of the billiard with various angles $\theta_0$ and compute the time-averaged projection $\bar{Q}$ both in space and in phase space. The quantum trails in the eigenstates result in an enhancement of $\bar{Q}$ along the short-time trajectory through $(\bm{q}_0, \bm{k}_0)$ (dashed white line). The system thus retains a memory of its initial condition and breaks ergodicity. In real space, the enhancement of $\bar{Q}(\bm{q})$ is particularly evident in the form of ``caustics'' on the short-time trajectories (see~\cite{SM} for enlarged figures). For wave packets launched along unstable periodic orbits (UPOs, for $\theta = 0^\circ, 30^\circ$) the effect appears slightly stronger, which can be explained by quantum scars. Here, $k = 148.80$.}
		\label{Fig3}
	\end{figure*}
	
	We begin by considering in Fig.~\ref{Fig2}(b) a Haar random state~\footnote{Haar random states are obtained in the space-basis upon drawing $\langle \bm{q}_n | \psi_{\rm Haar} \rangle = A(\alpha_n + i \beta_n)$, with $\alpha_n$ and $\beta_n$ normally distributed random numbers, $A$ a normalization constant, and $\bm{q}_n$ the discretized space points. This definition depends on the space discretization step, but the projection $Q_{\psi}(\bm{q}\bm{k})$ shown in Fig.~\ref{Fig2}(b) will be qualitatively the same as long as the discretization step is much smaller than $\frac{2\pi}{k}$, which is the case in our computations.}. The space projection $Q_{\rm Haar}(\bm{q})$ is a collection of random pixels, and the phase-space projection $Q_{\rm Haar}(\bm{q}\bm{k})$ exhibits a speckled pattern analogous to that predicted in Fig.~\ref{Fig1}(a). A closer analog of the eigenstates is a random wave superposition with momentum $k$~\cite{berry1977regular}, namely, $\langle \bm{q} |  \psi_{\rm rw} \rangle \sim  \sum_{n = 1}^N e^{i (\bm{k}_n \cdot \bm{q} + \phi_n)}$ (see~\footnote{More precisely, we obtain the random wave superpositions as follows. In position basis we consider a sum $\sum_{n = 1}^N e^{i \bm{q} \cdot \bm{k}_n + \phi_n}$ of waves with momentum $\bm{k}_n = k (\cos \theta_n, \sin \theta_n)$, and $\theta_n$ and phase $\phi_n$ uniform random numbers in $[0,2\pi]$. In an effort to make the random wave superposition as close as possible to actual energy eigenstates, we symmetrize it such that $\psi_{\rm rw}(x,y) = \pm \psi_{\rm rw}(x,-y) = \pm \psi_{\rm rw}(-x,y)$, and make it vanish on the boundary and outside of it by multiplying it by a factor $e^{-k \ell(\bm{q})}$, with $\ell(\bm{q})$ the distance of $\bm{q}$ from the boundary of the billiard and $\ell(\bm{q}) = \infty$ for $\bm{q}$ outside of the billiard. Finally, we renormalize so that $\langle \psi_{\rm rw} | \psi_{\rm rw} \rangle = 1$.} for details). In space, $Q_{\rm rw}(\bm{q})$ consists of a pattern with spatial features of size $\sim \frac{2\pi}{k}$, see Fig.~\ref{Fig2}(c). In phase space, $Q_{\rm rw}(\bm{q}\bm{k})$ is approximately uniform when moving along a trajectory, but only until the next collision with the boundary of the billiard, creating a \textit{partial quantum trail}. This is understood by considering three points $z,z^\prime$, and $z^\pprime$ on a trajectory, with no boundary collision separating $z$ and $z^\prime$ and with one boundary collision separating $z^\prime$ and $z^\pprime$, namely, $\bm{k} = \bm{k^\prime} \neq \bm{k}^\pprime$. The wave components that most contribute to $Q_{\rm rw}(z)$ and $Q_{\rm rw}(z^\prime)$ are the same, namely, those with $\bm{k}_n \approx \bm{k} = \bm{k}^\prime$, whereas the wave components that matter for $Q_{\rm rw}(z^\pprime)$ are different, namely, those with $\bm{k}_n \approx \bm{k}^\pprime$. That is, a trail connects $z$ and $z^\prime$, but not $z^\prime$ and $z^\pprime$.
	
	Finally, in Fig.~\ref{Fig2}(d) we consider the most interesting case of an energy eigenstate $\ket{k_n}$. The space projection $Q_{k_n}(\bm{q})$ is qualitatively similar to that of a random wave superposition~\cite{berry1977regular}. The phase-space projection $Q_{k_n}(\bm{q}\bm{k})$ is markedly different: it correlates radially along the trajectories, even when these go through collisions with the boundary of the billiard. That is, for the eigenstates we find and visualize full-fledged quantum trails. This is not contingent on the specific choice of the eigenstate: in Fig.~\ref{Fig2}(e) we show the quantum trails in $7$ consecutive eigenstates. The symmetries of the problem allow us to focus on just one quarter of the phase space. Some eigenstates appear clearly scarred, namely, localized along unstable periodic orbits (e.g., for $n = 5005$ and $5009$). Other eigenstates are not visibly scarred, and yet they exhibit quantum trails across the whole phase space (e.g., for $n = 5008, 5011$, and $5016$).
	
	A quantitative analysis is provided in Fig.~\ref{Fig2}(f) by computing the Pearson correlation coefficient $C_{\psi}$ between $Q_\psi(z_0)$ and $Q_\psi(z_t)$ for a fixed $\ket{\psi}$ and with respect to  an ensemble of pairs $(z_0,z_t)$. The latter are obtained sampling $z_0$ uniformly in phase space, and $z_t$ uniformly along the short-time trajectory from $z_0$, that is, sampling $t$ uniformly from $[0,\frac{3}{\lambda}]$, with $\lambda \approx 0.86 k$ the numerically computed Lyapunov exponent. The correlation $C_\psi$ is plotted versus the phase-space distance $d = |z_0 - z_t|$, which we naturally define as $d^2 \eqdef - \log(\left| \langle z_0 | z_t \rangle \right|^2) = \frac{k}{2}\left( |\bm{q}_0 - \bm{q}_t|^2 + \frac{|\bm{k}_0 - \bm{k}_t|^2}{k^2}\right)$. For the Haar random state, the correlation $C_\psi$ trivially relies on an overlap between $\ket{z_0}$ and $\ket{z_t}$, and thus quickly decays as $C_\psi \sim e^{-d^2}$. A longer correlation is found for the random wave superpositions, thanks to their partial trails. But it is the eigenstates that, thanks to their full-fledged trails, yield the strongest and longest correlations.
	
	Next, to see the memory effects that the quantum trails underpin, we launch a wave packet with position $\bm{q}_0 = (0,0)$ and momentum $\bm{k}_0 = k(\cos \theta_0, \sin \theta_0)$ and inquire about its time-averaged projection in Eq.~\eqref{eq. Qbar}, namely, $\bar{Q}(\bm{q} \bm{k}) = \sum_E Q_E(\bm{q} \bm{k}) Q_E(\bm{q}_0 \bm{k}_0)$, shown in Fig.~\ref{Fig3} for various initial directions $\theta_0$. The classical billiard is ergodic~\cite{bunimovich1974ergodic, bunimovich1979ergodic}, which suggests a uniform $\bar{Q}$ irrespective of $\theta_0$. By striking contrast, the quantum trails in the eigenstates imply that $\bar{Q}$ is enhanced along the short-time trajectory from $(\bm{q}_0,\bm{k}_0)$, which depends on $\theta_0$: the system retains memory of the initial condition and ergodicity is broken. An analysis of the timescales at which ergodicity breaking manifests, together with an explicit comparison with the classical (ergodic) case, is presented in~\cite{SM}.
	
	Our intuition was built in phase space, but the effects persist in real space. To see this we modify the time-averaged phase-space projection in Eq.~\eqref{eq. Qbar} into a time-averaged space projection~\cite{heller1980quantum}, $\bar{Q}(\bm{q}) = \sum_E \left| \langle \bm{q} | E \rangle \right|^2 Q_E(\bm{q}_0 \bm{k}_0)$, that is nothing but the probability density of finding the particle in position $\bm{q}$ at a random time $t \ggg 1$. Loosely speaking, we can think of $\bar{Q}(\bm{q})$ as obtained integrating out the momentum from $\bar{Q}(\bm{q}\bm{k})$, which can reduce the memory effect without suppressing it. Indeed, in the bottom of Fig.~\eqref{Fig3} we observe that $\bar{Q}(\bm{q})$ is enhanced along the short-time trajectory, particularly in the form of caustics. The contrast of the enhancement in $\bar{Q}$ appears slightly larger when the system is launched along an unstable periodic orbit, namely, for $\theta_0 = 0^\circ$ and $30^\circ$ in Fig.~\ref{Fig3}, which can be explained by quantum scarring. Our key finding is that ergodicity is broken more in general, due to quantum trails, even for trajectories that are not on unstable periodic orbits. Note that $\left| \langle \bm{q} | E \rangle \right|^2$ is invariant upon horizontal or vertical mirroring, and so is $\bar{Q}(\bm{q})$. Enlarged figures for $\bar{Q}(\bm{q})$ are shown in~\cite{SM}.
	
	In conclusion, we have unveiled the structure of chaotic eigenstates in phase space and the memory effects that it implies. The core intuition for quantum scars is that if a wave packet goes around an unstable periodic orbit with limited dispersion, then some eigenstates will be localized along the orbit and a system initialized on it is more likely to remain there~\cite{kaplan1998linear}. Quantum trails modify and extend this intuition to all the trajectories, even nonperiodic ones: if the wave packet follows a trajectory $z_t$ with limited dispersion, meaning $e^{-i\hat{H}t}\ket{z_0} \approx \ket{z_t}$, then quantum trails are left in the eigenstates, and a system initialized in $\ket{z_0}$ has an enhanced probability of being found on the short-time trajectory through $z_0$, even at long times. In quantum scars, like in semiclassical localization~\cite{bohigas1993manifestations,kaplan2000short}, Anderson localization~\cite{anderson1958absence}, and many-body localization~\cite{pal2010many, abanin2019colloquium}, the memory effect can be attributed to the localization of the eigenstates. Crucially, we have shown that eigenstate localization is in fact not necessary for a memory effect: correlations of the eigenstates along the trajectories, i.e., quantum trails, suffice. The condition for this, namely, that an initially localized wave packet should not immediately disperse in phase space, is a generous one, suggesting that trails and memory effects should be recurring features of quantum systems, both chaotic and non, opening the way to much further research. Beyond assessing these effects in other single-particle systems, such as dispersing billiards, softened billiards, and quantum maps, a particularly timely question regards the implications of quantum trails on many-body quantum systems, in particular with respect to thermalization, ergodicity, and entanglement.
	
	\textit{Note added}: While completing this paper, the author became aware through discussions with Prof.~E.~Heller and his group of closely related work that then appeared in Ref.~\cite{graf2024birthmarks}, in which the memory effects in Fig.~\ref{Fig3} are dubbed ``birthmarks'' and both the stadium billiard and block-random matrix models are studied.
	
	\textbf{Acknowledgements.}
	The author acknowledges support by Trinity College Cambridge and discussions related to this work with A.~Buchleitner, C.~B.~Dag, B.~Evrard, E.~Heller, W.~W.~Ho, J.~Knolle, A.~Lamacraft, M.~McGinley, and L.~Sá.
	
	\bibliography{biblio_quantum_trails}
	
	\appendix
	\section{Appendix A}
	In this Appendix we estimate the contrast of the memory effect in the time-averaged projection $\bar{Q}$, namely Eq.~\eqref{eq. contrast}.
	
	Consider a binning of the energy, and say $\mathcal{E}_n$ the set of eigenvalues $E$ in the $n$-th bin, namely with $n\epsilon \le E < (n+1)\epsilon$, with $\epsilon$ the bin width. Consider that the bin width can be taken large compared to the mean energy spacing, but small compared to the involved energy scales (e.g., $\langle z | \hat{H} | z \rangle$), which is possible for semiclassical and many-body quantum chaotic systems. Let us denote $\langle \dots \rangle_n = \mathcal{N}_n^{-1} \sum_{E \in \mathcal{E}_n} (\dots)$ the average over the $n$-th bin, with $\mathcal{N}_n \gg 1$ the number of eigenvalues in it. The sum over the energies can be written as a sum over bins, namely $\sum_E (\dots) = \sum_n \mathcal{N}_n \langle \dots \rangle_n$, and so
	\begin{align}
		\bar{Q}(z|z_0)
		= \sum_n & \mathcal{N}_n \langle Q_E(z) Q_E(z_0) \rangle_n, \\
		= \sum_n & \mathcal{N}_n \big[ \langle Q_E(z) \rangle_n \langle Q_E(z_0) \rangle_n + \dots \nonumber \\
		& \text{cov}_n \left( Q_E(z), Q_E(z_0) \right) \big], \\
		= \sum_n & \mathcal{N}_n \langle Q_E(z) \rangle_n \langle Q_E(z_0) \rangle_n \times \dots \nonumber \\
		& \left[1 + \text{cov}_n \left( \frac{Q_E(z)}{\langle Q_E(z) \rangle_n}, \frac{Q_E(z_0)}{\langle Q_E(z_0) \rangle_n} \right) \right],
	\end{align}
	where $\text{cov}_n$ denotes the covariance computed with respect to the ensemble of eigenvalues $E \in \mathcal{E}_n$. Let us assume that $\langle \bar{Q}_E^2(z) \rangle_n \approx 2 \langle \bar{Q}_E(z) \rangle_n^2$. This assumption is verified, e.g., if the normalized projections $\frac{\bar{Q}_E(z)}{\langle \bar{Q}_E(z) \rangle_n}$ have a Porter-Thomas (that is, exponential) distribution for $E \in \mathcal{E}_n$, which is a general universal feature of quantum chaotic systems~\cite{porter1956fluctuations,boixo2018characterizing}. Under this assumption, the variance of $\frac{\bar{Q}_E(z)}{\langle \bar{Q}_E(z) \rangle_n}$ is $1$, and the covariance can be replaced by a Pearson correlation coefficient $\text{corr}_n$, which has the advantage of taking the simple values $0$ and $1$ for uncorrelated and perfectly correlated variables, respectively. Because the Pearson correlation coefficient is invariant under scaling of its arguments, we can lift the denominators and get
	\begin{align}
		\bar{Q}(z|z_0)
		\approx \sum_n
		& \mathcal{N}_n \langle Q_E(z) \rangle_n \langle Q_E(z_0) \rangle_n \times \dots \nonumber \\
		& \left[1 + \text{corr}_n \left( Q_E(z), Q_E(z_0) \right) \right].
	\end{align}
	For a generic phase-space point $z_g$ not on the short-time trajectory from $z_0$, we have $\text{corr}_n \left( Q_E(z_g), Q_E(z_0) \right) \approx 0$. Instead, the quantum trails in the eigenstates mean that for a point $z_t$ on the short-time dynamics through $z_0$ the projections $Q_E(z_t)$ and $Q_E(z_0)$ are correlated, to an extent which we now estimate.
	
	The pointer state $\ket{z_t}$ fails to exactly follow the true dynamics $e^{-i\hat{H}t} \ket{z_0}$. We can thus write $\ket{z_t} = \sqrt{\alpha} e^{-i\hat{H}t} \ket{z_0} + \sqrt{1 - \alpha} \ket{r_t}$, with $\ket{r_t}$ a state orthogonal to $e^{-i\hat{H}t} \ket{z_0}$ and $\alpha = \left| \langle z_t | e^{-i \hat{H} t} | z_0 \rangle \right|^2$. We have
	\begin{equation}
		Q_E(z_t)
		= \left| \langle E | z_t \rangle \right|^2
		= \left| \sqrt{\alpha} e^{iEt} \langle E | z_0 \rangle + \sqrt{1-\alpha}\langle E | r_t \rangle \right|^2.
		\label{eq. Qezt}
	\end{equation}
	The state $\ket{r_t}$ by construction contains the bits of the wavepacket $e^{-i\hat{H}t} \ket{z_0}$ that get dispersed beyond $\ket{z_t}$, and it is thus sensible to assume that $\langle E | z_0 \rangle$ and $\langle E | r_t \rangle$ are statistically independent. Moreover, if the eigenstates in $\mathcal{E}_n$ are similarly spread across phase space due to chaos, then we can assume that $\langle Q_E(z) \rangle_n \approx \langle Q_E(z_0) \rangle_n$ for all the phase space points of interest $z$. Under these assumptions, from Eq.~\ref{eq. Qezt} we compute $\text{corr}_n \left(Q_E(z_t), Q_E(z_0)\right) \approx \alpha$, and thus get
	\begin{align}
		\bar{Q}(z_t|z_0)
		& \approx \left[1 + \left| \langle z_t | e^{-i \hat{H} t} | z_0 \rangle \right|^2 \right] \sum_n \mathcal{N}_n \langle Q_E(z_0) \rangle_n^2, \\
		\bar{Q}(z_g|z_0)
		& \approx \sum_n \mathcal{N}_n \langle Q_E(z_0) \rangle_n^2.
	\end{align}
	The contrast in the trail of the time averaged projection $\bar{Q}(z|z_0)$ thus yields
	\begin{equation}
		\frac{\bar{Q}(z_t|z_0)}{\bar{Q}(z_g|z_0)}
		\approx 1 + \left| \langle z_t | e^{-i \hat{H} t} | z_0 \rangle \right|^2.
	\end{equation}

	\clearpage
	
	\setcounter{equation}{0}
	\setcounter{figure}{0}
	\setcounter{page}{1}
	\thispagestyle{empty} 
	\makeatletter 
	\renewcommand{\figurename}{Fig.}
	\renewcommand{\thefigure}{S\arabic{figure}}
	\renewcommand{\theequation}{S\arabic{equation}}
	\setlength\parindent{10pt}
	
	\onecolumngrid
	
	\begin{center}
		{\fontsize{12}{12}\selectfont
			\textbf{Supplementary Material for\\``Quantum trails and memory effects in the phase space of chaotic quantum systems"\\[5mm]}}
		{\normalsize Andrea Pizzi \\[1mm]}
	\end{center}
	\normalsize
	
	This Supplementary Material is devoted to a few details on the stadium billiard. In Section I we elaborate on the methods for the numerics, in Section II we provide high resolution zooms of the time-averaged wavefunction projection $\bar{Q}(\bm{q})$, and in section III we emphasize the importance of ``quantumness'' by directly comparing the quantum and classical cases.
	
	\section{I - Details on numerics}
	We solve the eigenvalue problem for the stadium billiard using the boundary integral method and following closely Ref.~\cite{backer2003numerical}. This method recasts the eigenproblem $\hat{H} \ket{k_n} = \frac{k_n^2}{2} \ket{k_n}$ to one on the boundary of the billiard. For a billiard with radius $R = 1$ and width equal to twice the height, the total perimeter is $\mathcal{L} = 2\pi + 4$. The space discretization step is $\Delta = \frac{\mathcal{L}}{M}$, where $M$ is the number of boundary discretization points. The discretization $\Delta$ sets an upper bound to the eigenvalues $k_n$ that we can resolve: we again follow Ref.~\cite{backer2003numerical} and set $k_{\rm max} = \frac{2\pi}{10 \Delta}$. We say $N$ the number of eigenstates with $k_n < k_{\rm max}$.
	
	A wavepacket $\ket{\bm{q}\bm{k}}$ has position standard deviation $\frac{\sigma}{\sqrt{2}}$ and momentum standard deviation $\frac{1}{\sqrt{2} \sigma}$, where $\sigma = \sigma(k) = \frac{1}{\sqrt{k}}$ is chosen to guarantee a good localization in phase space (see main text). For a wavepacket to be well-represented in terms of the numerically available eigenstates, we require that up to 4 standard deviations of the wavepacket should be within the resolved $k_{\rm max}$, that is, that $k + \frac{4}{\sqrt{2} \sigma(k)} < k_{\rm max}$. This condition can be massaged into $k < (\sqrt{k_{\rm max}+2}-\sqrt{2})^2$. In the main text, we consider $M = 2000$ in Fig.~2, for which $N = 8386$, $k_{\rm max} \approx 122$, $k \approx 94.67$, and $\sigma \approx 0.103$, and $M = 3000$ in Fig.~3, for which $N = 18946$, $k_{\rm max} \approx 183.30$, $k \approx 148.80$, and $\sigma \approx 0.082$. In Fig.~2 we consider the eigenstates with $k_n \approx k$, namely $k_{5005} \approx 94.544$, $k_{5006} \approx 94.552$, $k_{5007} \approx 94.583$, $k_{5008} \approx 94.593$, $k_{5009} \approx 94.599$, $k_{5010} \approx 94.603$, $k_{5011} \approx 94.639$, $k_{5016} \approx 94.660$.
	
	\begin{figure}
		\centering
		\includegraphics[width=\linewidth]{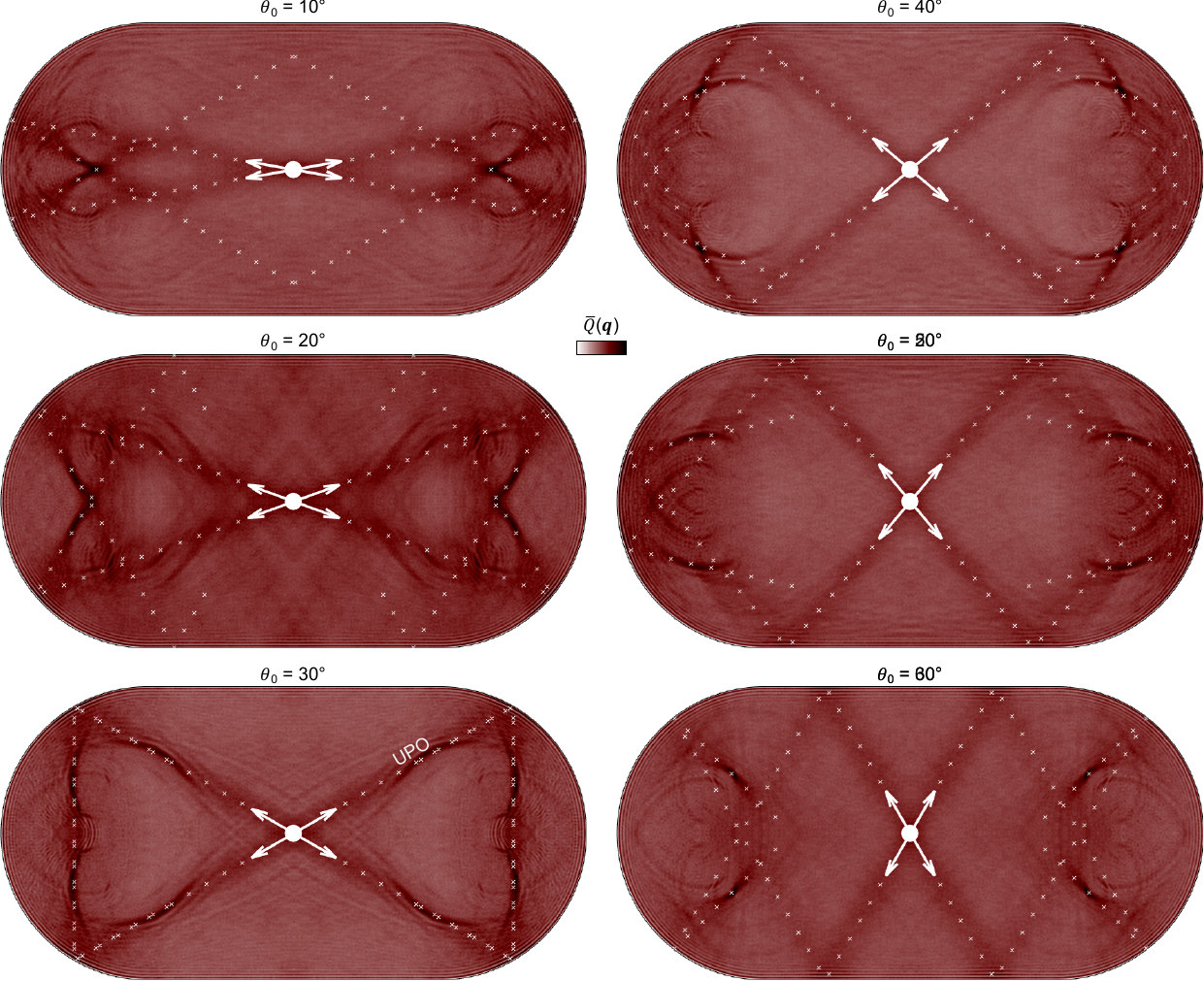}
		\caption{\textbf{Details of the time-averaged projection.} We report the results for $\bar{Q}(\bm{q})$ from Fig.~3 in the main text, but with larger resolution and overlapping not just the short-time classical trajectory, but also its mirrored copies. The long-time projection of the wavefunction is enhanced along the short-time classical trajectory. It can be fully appreciated that the caustics of $\bar{Q}(\bm{q})$ tend to lie on the short-time trajectories, that depend on the initial angle $\theta_0$.}
		\label{FigS1}
	\end{figure}
	
	\section{II - High resolution time averaged projections}
	In Fig.~\ref{FigS1} we report the same plots of  $\bar{Q}(\bm{q}) = \sum_E \left| \langle \bm{q} | E \rangle \right|^2 Q_E(\bm{q}_0 \bm{k}_0)$ as in Fig.~3 in the main text, but enlarged to improve visibility. Moreover, because all eigenstates are either symmetric or anti-symmetric upon vertical and horizontal reflections, the projection $\left| \langle \bm{q} | E \rangle \right|^2$ is symmetric with respect to such reflections, and so is the time-averaged projection $\bar{Q}(\bm{q})$. In order to understand the structure in $\bar{Q}(\bm{q})$, it is thus sensible to overlap not only the classical short-time trajectories starting from $(\bm{q}_0 \bm{k}_0)$, but also its three mirrored versions obtained upon horizontal reflection, vertical reflection, or both. This allows to fully appreciate how much of the structure of $\bar{Q}(\bm{q})$, and in particular its caustics, can be predicted from the short-time classical trajectories.
	
	\section{III - Role of quantumness and timescale for ergodicity breaking}
\begin{figure}
		\centering
		\includegraphics[width=\linewidth]{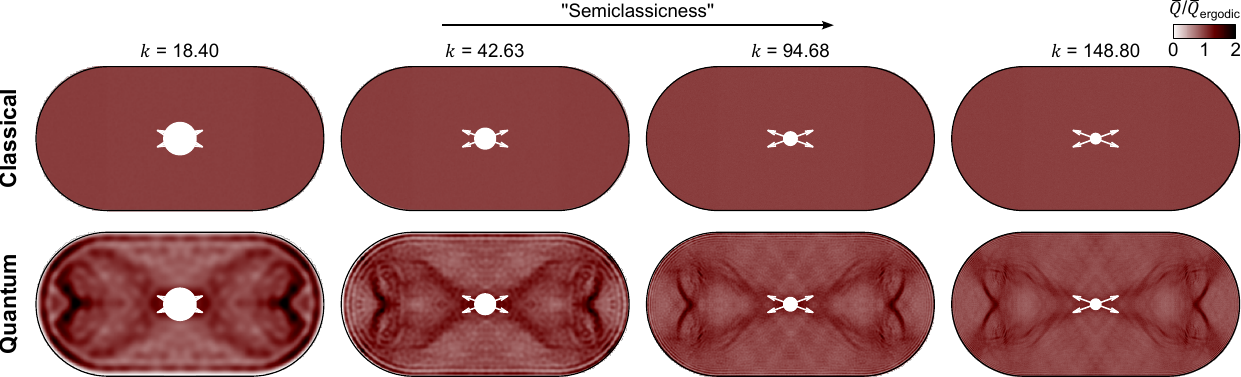}
		\caption{\textbf{Time-averaged distributions: classical vs quantum.} We compare the classical and quantum time-averaged space distributions. 
			Classically (top), we obtain a uniform distribution, due to ergodicity~\cite{bunimovich1974ergodic, bunimovich1979ergodic}. Quantum mechanically (bottom), we instead observe the memory effects described in the main text. Going deeper into the semiclassical limit (increasing $k$ left to right) yields narrower trails, and thus narrower features of the quantum distribution (a white circle with radius $\sigma = \frac{1}{\sqrt{k}}$ shows the relevant length scale, corresponding to the standard deviation of the initial condition). The initial condition is a wavepacket launched from the middle of the billiard with angle $\theta_0 = 20^\circ$. An arrow pointing along the initial wavepacket direction is shown in white, together with its 3 symmetric copies (as done for Fig.~\ref{FigS1}). The classical time-averaged distribution $\bar{Q}_c(\bm{q})$ was obtained from $R = 10^6$ trajectories and averaging over times $500<t<1000$.}
		\label{FigS2}
	\end{figure}

	We emphasize how our results rely on quantum mechanics by comparing them with classical simulations. Quantum mechanically, we consider the dynamics starting from a Gaussian wavepacket $\ket{\bm{q}_0\bm{k}_0}$, as in Fig.~3, and compute the quantum probability distribution $Q_q(\bm{q},t) = \left| \langle \bm{q} | e^{-i\hat{H}t} | \bm{q}_0\bm{k}_0 \rangle \right|^2$. Classically, we consider an ensemble of trajectories with initial positions and momenta distributed according to a Gaussian with mean and variance matching those of $\ket{\bm{q}_0\bm{k}_0}$, and use them to reconstruct the dynamics of a probability distribution $Q_c(\bm{q},t)$. In Fig.~\ref{FigS2} we consider the time-average of such distributions,
	\begin{equation}
		\bar{Q}_{c,q}(\bm{q}) = \lim_{t \to \infty} \frac{1}{t} \int_0^t d \tau Q_{c,q}(\bm{q},t),
		\label{eq. Q bar}
	\end{equation}
	for various values of the effective inverse Planck constant, $k$, which controls the ``semiclassicness'' of the system. The classical distribution $\bar{Q}_c$ is uniform, irrespective of $k$ and due to ergodicity, that for the classical stadium billiard has been proven~\cite{bunimovich1974ergodic, bunimovich1979ergodic}. As discussed in the main test, the quantum distribution $\bar{Q}_q$ displays instead a memory of the initial condition. Thus, while the classical wavepacket gets fully mixed by the dynamics and leads to a uniform distribution, the quantum wavepacket does not -- ergodicity is broken due to a genuinely quantum interference effect. The larger $k$, the smaller the size $\sigma = \frac{1}{\sqrt{k}}$ of the initial wavepacket, the narrower the quantum trails, and the narrower the regions of enhanced probability in $\bar{Q}_q$. For $k \to \infty$, the trails become infinitely narrow and ergodicity is recovered.
	
\begin{figure}
		\centering
		\includegraphics[width=\linewidth]{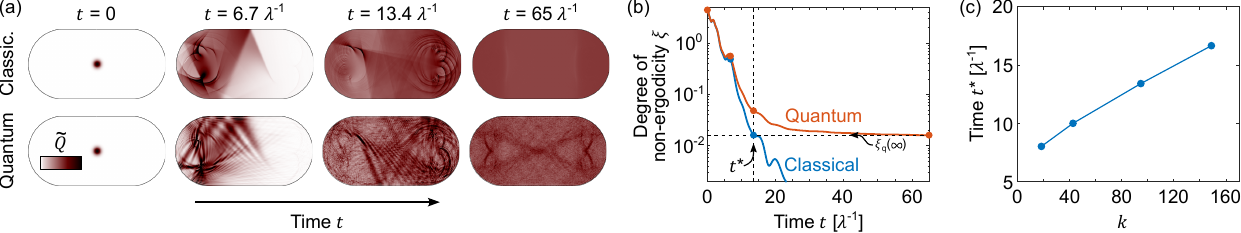}
		\caption{\textbf{Timescale of ergodicity breaking.} (a) Classical and quantum dynamics of $\tilde{Q}(t)$, namely the probability distribution averaged over a moving time window, see Eq.~\eqref{eq. Q tilde}. At short times the two are in good correspondence. Eventually, the classical distribution becomes uniform due to ergodicity, whereas the quantum one maintains a memory of the initial condition. (b) The non-ergodicity of the system is measured by $\xi$, the Kullback–Leibler divergence between $\tilde{Q}$ and the uniform distribution (shown for $k = 94.68$). In the quantum case, ergodicity is broken and $\xi_q$ tends to a finite value $\xi_q(\infty)$ (horizontal dashed line). In the classical case, ergodicity occurs and $\xi_c \to 0$. We call $t^*$ the time at which $\xi_c(t^*) = \xi_q(\infty)$ (vertical dashed line), namely the time at which ergodicity breaking becomes manifest. The times considered in (a) are marked by dots. (c) The time $t^*$ is shown versus the effective inverse Planck constant $k$, suggesting a proportionality relation. Times are shown in units of the inverse Lyapunov exponent $\lambda^{-1}$. Here, the initial wavepacket is launched from the center of the billiard with an angle $\theta_0 = 20^\circ$.}
		\label{FigS3}
	\end{figure}

	Having shown the difference between the classical and quantum time-averaged distributions, we now ask at what time can such difference be appreciated. To this end, we introduce a partially time-averaged distribution,
	\begin{equation}
		\tilde{Q}_{c,q}(\bm{q},t) = \frac{2}{t} \int_{t - \frac{t}{4}}^{t + \frac{t}{4}} d\tau \ Q_{c,q}(\bm{q},\tau),
		\label{eq. Q tilde}
	\end{equation}
	which averages the probability distribution over a time interval $(t - t/4,t + t/4)$ that both shifts and expands with $t$. The distribution with a tilde in Eq.~\ref{eq. Q tilde} reaches that with a bar in Eq.~\ref{eq. Q bar} at infinite times, namely $\tilde{Q}_{c,q} \xrightarrow{t \to \infty} \bar{Q}_{c,q}$. The classical and quantum distributions $\tilde{Q}_{c,q}$ are shown for various times in Fig.~\ref{FigS3}(a). At short times they are similar. But as soon as enough mixing has taken place, a key difference becomes manifest: the classical distribution becomes more uniform than the quantum one, that instead retains a memory of the initial condition. In Fig.~\ref{FigS3}(b) we quantify the degree of non-ergodicity as $\xi_{c,q}(t) = D_{KL}(\tilde{Q}_{c,q}(t) || Q_{\rm erg})$, namely as the Kullback–Leibler divergence between $\tilde{Q}_{c,q}(t)$ and an a ergodic distribution $Q_{\rm erg}$, that is uniform inside the billiard and zero elsewhere. Classically, ergodicity entails that $\tilde{Q}_c(t)$ tends to $Q_{\rm erg}$, and thus $\xi_c \xrightarrow{t \to \infty} 0$. Quantum mechanically, $\tilde{Q}_q(t)$ fails to reach $Q_{\rm erg}$, and thus $\xi_q \xrightarrow{t \to \infty} D_{KL}(\bar{Q}_{q} || Q_{\rm erg} ) > 0$ remains finite. We can thus define the time $t^*$ at which ergodicity breaking manifests as the time at which $\xi_c(t^*) = \xi_q(\infty)$, namely at which the classical distribution becomes more uniform than the quantum system will ever be. Such timescale depends on how quickly the wavepacket spreads in phase space, and is thus arguably related to the Thouless time. We show $t^*$ in Fig.~\ref{FigS3}(c). Intuitively, $t^*$ grows with $k$: the larger $k$, and the longer it takes for the classical system to become more ergodic than the quantum one, both because the classical system is initially more localized and because the quantum system has narrower traces in $\bar{Q}_q$, thus a lower $\xi_q(\infty)$.
	
\end{document}